\documentclass[preprint,amsmath,amssymb,superscriptaddress,showpacs]{revtex4}
\usepackage{graphicx}
\begin{document}
\baselineskip=0.8cm
\renewcommand{\thesection}{\arabic{section}}
\renewcommand{\thesubsection}{\arabic{section}.\arabic{subsection}}
\renewcommand{\thefigure}{\arabic{figure}}
\title{Ground state properties of a confined simple atom by C$_{60}$ fullerene}
\author{M. Neek-Amal}
\affiliation{Department of Nano-Science,
 Institute for Studies in Theoretical Physics and Mathematics (IPM),
P.O. Box 19395- 5531,Tehran, Iran.}
\author{G. Tayebirad}
\affiliation{Department of Physics, Iran University of Science and
Technology, Tehran, Iran.}
\author{R. Asgari~\footnote{Corresponding author: asgari@theory.ipm.ac.ir}}
\affiliation{Institute for Studies in Theoretical Physics and
Mathematics, Tehran 19395-5531, Iran.}

\begin{abstract}
We numerically study the ground state properties of endohedrally
 confined hydrogen (H) or helium (He) atom by a molecule
of C$_{60}$. Our study is based on Diffusion Monte Carlo method. We
calculate the effects of centered and small off-centered H- or
He-atom on the ground state properties of the systems and describe
the variation of ground state energies due to the C$_{60}$
parameters and the confined atomic nuclei positions. Finally, we
calculate the electron distributions in $x-z$ plane in a wide range
of C$_{60}$ parameters.
\end{abstract}

\maketitle

\section{Introduction}
\label{sec1}

The Properties of Carbon allotropes with closed cage structures have
been an active subject of research since its first experimental
discovery back in 1985~\cite{kroto}. Accurate calculation of the
structural properties of single shell- or tubes fullerenes have been
performed by H\"{a}ser {\it et al.}\cite{haser} using the
Hartree-Fock methods, by Dunlap {\it et al.}\cite{dunlap} and Heggie
{\it et al.}\cite{heggie} using density functional theory. The later
group considered fullerenes up to a large number of atoms. The
unique spherically shaped cage of a C$_{60}$ fullerene contains
empty space that is large enough to incorporate atoms of any kind of
endohedral composite. The molecule is not very reactive since the
carbon atoms are all completely saturated by taking part in an
extended $\pi-$bonded electron system similar to graphite. The
interior of C$_{60}$ is expected to have a good capacity for
confining small molecules, ions and
atoms~\cite{shinohara,dolmatov,amusia1,connerade,c60}. Physical
properties of a confined simple atom by C$_{60}$ fullerene is a
subject of significant interest in recent years, mostly because they
exhibit properties that can lead to important applications in
nanostructure science and technology~\cite{nature}. Furthermore,
many researchers have been working on modeling the storage of
various types of gases and metal atom in nanotubes and
fullerenes~\cite{Hcnt,yufe}. This type of research is also
fundamental in the application of nano-technology to gas storage
devices, used in such systems as fuel cells, construction of
molecular sieves and filtration membranes.

In addition to the real atomic interaction approach, one can modeled
the interaction between adsorbed atoms to C$_{60}$ molecule by an
effective attractive potential which can hold electrons of atom near
C$_{60}$ wall. This effective potential, in effect, replaces the
C$_{60}$ walls and has been successful in explaining many
experimental and computational results \cite{Puska,Xu}. It is
physically expected to observe significant difference between the
ground state properties of the confined atoms by C$_{60}$ fullerene,
and their free states \cite{conn2}. Obviously, in this case the
strength of the attractive potential is not so strong as to
immobilize electrons. Recently, some authors~\cite{sen2,conn3} used
a similar model of C$_{60}$ molecule constructed by Xu {\it et
al.}~\cite{Xu} to calculate the properties of the endohedrally
confined atoms.

Researches using modeling studies in this field have employed
various computational methods based on prescribed inter-atomic
potential energy functions. The Schr\"odinger equation provides the
accepted description for microscopic phenomena at non-relativistic
energies. This equation can be solved analytically only in a few
highly idealized models. Calculation of the ground state properties
of a system with several atoms/molecules or with an arbitrary
boundary conditions is a scientific efforts. A theory based on
Quantum Monte Carlo (QMC) methods allows us to calculate the ground
state properties more accurately. For the many body system, QMC has
been very useful in providing exact results.

The main purpose of the present work is to calculate the effect of a
centered or small off-centered Hydrogen (H) or Helium (He) atom on
the ground state energy and describe also the variation of the
ground state energies on the strength of attractive model potential
within diffusion Monte Carlo (DMC) method. Although the simple
spherical shell model for C$_{60}$ molecule does not completely
describe the real endohedrally off-centered atoms, but we use it for
considering the off-centered effects as a first approximation.

The contents of the paper are briefly as follows. In section II, we
review the basic formalism, including our model to the system and in
the next section we briefly discuss the DMC method with the
appropriate steps ,as well as the employed trial wave function. Our
numerical results for the atoms confined by C$_{60}$ fullerene are
presented in Sec. IV. Finally, in Sec. V we summarize our main
conclusion and discuss our further works.

\section{Theory}
\label{sec2}

Our approach is based on modeling an endohedral medium by a
short-range spherical shell with an attractive potential $U_0$ in
the range $a \leq r\leq ka$, where $a$ and $ka$ are the inner and
outer radii of the shell with $\Delta=a(k-1)$ as the thickness of
the shell. Fig.~1 depicts all the information on our selected model
in the case of a H-atom confined endohedrally within a C$_{60}$
fullerene. In the case of He-atom we have a similar model and a
figure with two electrons can be fancied.

When an electron associated with an atom enters the shell medium, it
will be affected by further attractive potential, $U_0$, in addition
to the Columbic potentials due to other electrons and nuclei.
Therefore the potential energy is written as
\begin{equation}
V(\vec{r_i},\vec{x})=\left\{
\begin{array}{l}
\sum_{j\neq i} \frac{e^2}{\left|\vec{r_i}-\vec{r_j}\right|}-\frac{Ze^2}{\left|\vec{r_i}-\vec{x}\right|}+U_0,\,\hspace{0.3 cm}\mbox{for $a \leq r_i \leq ka $}\\
\sum_{j\neq i}
\frac{e^2}{\left|\vec{r_i}-\vec{r_j}\right|}-\frac{Ze^2}{\left|\vec{r_i}-\vec{x}\right|}\,\,\,\,\hspace{1.12
cm} \mbox{otherwise}~,
\end{array}
\right.
\end{equation}

where $\vec{x}$ is the position of nucleus, $\vec{r_i}$ represents
the position of $i$th electron which is measured from geometrical
center of the shell and $Z$ is the atomic number. Summation runs
over the number of electrons. The atomic nucleus is always assumed
to be fixed at position  $\vec{x}$. Moreover, we calculated the
effect of small off-centered nuclear position on ground state
properties by moving the position of nucleus along a direction and
assume it fixed in each simulation.

Xu {\it et al.}\cite{Xu} found a semi-empirical function for the
potential of C$_{60}$ fullerene and calculated the parameters $a$
and $b$ by fitting them to experimental data. Specifically, they
found $a\approx 5.75$ au which is approximately equal to the radius
of C$_{60}$ and $\Delta\approx 1.89$ au. The value of the shell
thickness, $\Delta$, is assumed to be different by different
authors~\cite{Puska}. We fixed  $a$ and $\Delta$ by the above values
in most parts of our calculations but in order to take the variation
of $\Delta$ into account, we also calculate the ground state energy
as a function of thickness by considering that for every simulation
these parameters are fixed. Although the effective value of the
attractive potential $U_0$ is used in Ref.~\cite{Amusia1} from
0.3015 to 0.4228 au, we examined our results in a wide range of
values for $U_0$. This is because of the number of carbon atoms
which can be changed in different systems.

We have used the well-known DMC method \cite{anderson,DMC,thez} to
calculate the ground state energies, $E$ of simple endohedrally
confined atoms such as H- or He-atom. The number of electrons for
such a system are less than three and thus the ground state wave
function has no node, consequently the Pauli exclusion principle
does not play a role. Note that it is necessary to use the
fixed-node QMC calculation\cite{DMC} or Exact Cancelation method for
a system with large number of fermions \cite{solidqmc}. In what
follows, we briefly discuss the DMC method used in our numerical
calculations.

Through an analytic continuation of time $t$ to
imaginary values $-i\tau$, the Schr\"{o}dinger equation of the system is
given by
\begin{eqnarray}
\frac{\partial\psi({\bf R},\tau)}{\partial\tau}= D\nabla^2_{{\bf r}}
\psi({\bf R},\tau) +(E_T-V({\bf R}))\psi({\bf R},\tau)~,
\label{eqn6}
\end{eqnarray}

where the constant $D=\hbar/ 2m$, $m$ is electron mass and $E_T$ is
an offset energy which can be adjusted at the beginning from
contributing an average potential for all the initial positions of
particles. In the limit $\tau \rightarrow \infty$, $\psi({\bf
R},\tau)$ tends to either a non-zero value or an infinite value
provided by $E_T$.

To avoid the effect of singularity in the potential, it would be
essential to use a guiding trial wave function. More specifically,
at the positions where $V({\bf R})$ diverges, the creation or
annihilation rate, $E_T-V({\bf R})$ becomes tremendously large and
makes the numerical algorithm unstable. In this case, we used a time
independent guide function $\psi_T({\bf R})$ by introducing a
function $\rho({\bf R} ,\tau)$ giving by $\rho({\bf
R},\tau)=\psi_T({\bf R})\psi({\bf R},\tau)$. By substituting
$\rho({\bf R},\tau)$ in Eq.~(\ref{eqn6}) we have
\begin{eqnarray}
\frac{\partial\rho({\bf R},\tau)}{\partial\tau}&=& D
\nabla_{{\bf R}}[ \nabla_{{\bf R}} -F({\bf R}) ]\rho({\bf R},\tau)\nonumber\\
&-&(E_L({\bf R})- E_T)\rho({\bf R},\tau)~, \label{eqn7}
\end{eqnarray}
where $F({\bf R})=2\nabla\psi_T({\bf R})/ \psi_T({\bf R})$ is the
so-called {\it Fokker-Planck} force and corresponds to a force
function which drifts or walks away from regions where
${|\psi_T({\bf R})|}^2$ becomes small. Meanwhile, the local energy
function is $E_L(\bf R)$$=-D \nabla_{{\bf R}}^2 \psi_T({\bf R})/
\psi_T({\bf R}) +V({\bf R}).$ It is important to choose the guide
function such that $E_L(\bf R)$ does not have any singularity. We
used the Green function method to solve Eq.~(\ref{eqn7}). We follow
the standard way to decompose the Green function into two parts. The
usual diffusive part is given by \cite{solidqmc}
\begin{eqnarray}
G_D({\bf R},{\bf R'},\Delta \tau)=\frac{exp(-\frac{({\bf R'}-{\bf
R}- D F({\bf R})\Delta\tau)^2}{4
D\Delta\tau})}{(4\pi~D~\Delta\tau)^{\frac{3N}{2}}} +o(\Delta
\tau)^2~, \label{eqn8}
\end{eqnarray}
 and the branching term which is given by
\begin{eqnarray}
G_{B}({\bf R}, {\bf R'}, \Delta\tau)=\exp[-\Delta
\tau(\frac{E_L({\bf R})+E_L({\bf
R}^{\prime})}{2}-E_T)]~.\label{eqnb1}
\end{eqnarray}
In order to reduce numerical errors to order $ (\Delta\tau)^2$, we
have used the Metropolis procedure.
 After performing the diffusion and branching processes for all the
particles, the final MC step lies in updating the value of $E_T$
on an ensemble average.
\begin{eqnarray}
E_T = <E_L> -\frac{N-N_0}{N_0~\Delta\tau }~, \label{eqn11}
\end{eqnarray}
where $N_{0}$ refers to the initial number of walkers and $N$
denotes number of walkers updated. This sort of adapting $E_T$ is
essential to avoid large amount of fluctuations in the number of
walkers. The ground state of the system is obtained by averaging
$E_T$ over many MC steps. Prior to presenting our results, it would
be illustrative to discuss the numerical errors in the DMC method.
Principally, there are two types of errors which limits the accuracy
of most DMC calculations: (a) Statistical or sampling errors
associated with the limited number of independent sample energies
used in determining the ground state energy. (b) The systematic
errors associated with finite time-step $\Delta \tau$, round-off in
computing, imperfectness of random number generators etc. The total
errors in our numerical calculation is about $10^{-4}$ for more or
less all cases.

\section{Numerical Results}

We started by focusing on the centered H-atom. Our choice of radial
part of ground state wave function is the linear composition of an
unperturbed H-atom wave function and outer well wave function. For
having the unique function with respect to $U_0$, we calculated the
radial ground state wave function by creating the same results as in
Ref. \cite{conn2} for several values of $U_0$ and then we found the
best fit with some constants depending to $U_0$ value. Our trial
wave function for every $U_0$ value is
\begin{eqnarray}
\psi_H(r)=A\exp(-B r)+C\exp(-E{(r-r_0)}^2)~, \label{eqn10}
\end{eqnarray}
where $r_0$ is approximated by the shell's mean point and is about
$a+\Delta/2$.

When $U_0$ is close to zero, we have a free electron ground state
properties and when $U_0$ is large enough, we have square cosine
function for a wave function. By using the cusp condition which
leads to $\psi_H '(r)+\psi_H(r)=0$, we can determine parameter $B$
in terms of other parameters as
\begin{eqnarray}
B=\frac{\psi_H(0)+2CEr_0\exp(-Er_0^2)}{A}~. \label{eqn10-b}
\end{eqnarray}
This condition guarantees to cancel singularity at small $r-$
region. From the fitting process mentioned above, we obtained
\begin{eqnarray}
\left\{
\begin{array}{ll}
A=3.06\exp(2.99~U_0) \\
C=\frac{0.146~{U_0}^2-0.03~U_0}{{U_0}^2-0.58~U_0+0.33} \\
E=\frac{1.34~{U_0}^2-0.17~U_0}{{U_0}^2-2.02~U_0+0.81}
\end{array}
\right. \label{eqn11}
\end{eqnarray}

For the He case in our system, we build a suitable trial function which is a
product of two simple H-atom trial wave function connected by a
Pad$\acute{e}$-Jastrow function
\begin{eqnarray}
\psi_{He}(r_1,r_2)=\psi_H(r_1)\psi_H(r_2) \varphi(r_{12})
,\label{eqn12}
\end{eqnarray}
where $r_{12}=|\vec{r_1}-\vec{r_2}|$,
$\varphi(r_{12})=\exp(r_{12}/2(1+ \alpha r_{12}))$ and $\alpha=0.2$
is a well-known value for He system\cite{joslin}. The parameter $B$
is fixed by using the cusp condition for the He case
\begin{eqnarray}
B=\frac{2\psi_H(0)+2CEr_0\exp(-Er_0^2)}{A}~.
\end{eqnarray}
Note that the pre-factor 2 in the first term of the nominator comes
from the fact that there are two electrons in the system with
respect to the H system.

To Consider small values of off-center( $\delta < a$) for both H and
He cases, we have assumed that the wave function of the system is
again given by Eq.~(\ref{eqn10}) and Eq.~(\ref{eqn12}) and one only
needs to shift the $r$ parameter by a small amount so that
$r\rightarrow \left|\vec{r}+ \vec{\delta} \right |$ where $\delta$
is small. Furthermore we restrict our calculations to a range such
that $\delta$ becomes less than $0.5a$ along the z-axis.

In our simulations we fixed $\Delta \tau\approx 0.05$. ( See the
appendix for more details). Throughout the paper, we have
implemented the following atomic units. Here and hereafter we use
the Bohr radius, $\hbar^2/me^2=0.529$ {\AA} as the length unit, the
value of $\hbar^3/me^4=2.419\times 10^{-2}$ fs as time unit and the
Hartree, $me^4/\hbar^2=27.25$ eV as the energy unit. The main
results of our work are shown in Figures 2-7.

\subsection{H-atom endohedrally confined in C$_{60}$ molecule}

First of all we calculated the ground state energy of H-atom
endohedrally confined within C$_{60}$ molecule as a function of
$U_0$, the confining well strength for different small nuclei
positions. The nucleus vector position is considered as
$\vec{x}=(0,0,Da)$ where $D$ assumes different values of $0, 0.25$
and $0.5$. For off-centered cases, H-atom obviously does not have
any spherical symmetry and as expected from an analytical approach,
it becomes either more difficult or undoable for such
configurations. It is apparently obvious from Fig.~2a that for the
small $U_0$-values, the ground state energy remains constant at the
usual well-known value of -0.5 for ground state energy (-13.6 eV).
It means that the electron is not influenced by the outer confining
well. For the case of $D=0$, our results are in excellent agreement
with recent exact results which were obtained by solving the
Schr\"odinger equation directly\cite{conn2}.
 From this figure, when the H-atom moves towards to the confining well region,
the energy maintains it's -0.5 value. It then starts to decrease,
when it reaches the threshold point at $U_0\simeq -0.7$. The value
of $U_0$ at the threshold point depends on the assumed values of
$D$. As physically expected by moving the nucleus of the H-atom far
from the shell, the threshold value of $U_0$ should move to the
right because the atom approaches the free atom case. The effect of
thickness for the ground state energy is shown in Fig.~1b with
$a=5.75$. The electron of the confined H-atom becomes more immobile
when the shell thickness increases. Consequently the electron
becomes localized by increasing the shell thickness or the value of
the attractive potential. For sake of clarification, the numerical
data of the ground state energies are given in the Tables I and II.

The probability of finding the electron at a distance $r$ from its
nucleus is proportional to $r^2\rho(r)$ where
$\rho(r)=\phi(r)\psi_H(r)$ and $\phi(r)$ is the electron radial wave
function assuming that the nucleus of the H-atom remains at the
origin. Fig.~3 depicts the $r\phi(r)$ as a function of $r$ for
different $U_0$-values. The electron does not feel the effect of the
well for small $U_0$-values, however it is trapped in the shell
region for large $U_0$-values. It is important to note that the
$r\phi(r)$ does not have any tail for very large attractive
potential.

When the H-atom is off-centered, we calculated two-dimensional
electron distributions in the $x-z$ plane. Fig.~4 shows the effects
of centered $(D=0 )$ and off-centered $(D=0.25)$ H-atom on the
electron distributions at threshold value of $U_0$. In the
symmetrical case, electron distribution is uniformly distributed.
Our results show that electron distribution has a peak around the
nucleus in all directions for a weak attractive potential. However,
the electron is localized in the vicinity of the shell region for a
strong attractive potential. In this case, the atom becomes ionized
depending on the large value of $U_0$ and $D$. As it is physically
understandable, the electron distributions along $x-$ or $y$-axis
holds its symmetry. However the electron distribution along $z$-axis
has no symmetry.

\subsection{He-atom endohedrally confined in C$_{60}$ molecule}

To the best of our knowledge, there is no analytical solution for
He-atom confined within C$_{60}$ molecules. Recently, some authors
have intensively studied the two-electron photoionization cross
section of He atom confined in C$_{60}$.\cite{amusia1} They
calculated the regular and irregular solutions of the
Schr\"{o}dinger equation for the system. Here we present more
accurate ground state energy as well as the associate wave function
for the system which will be used to calculate other physical
quantities.

We considered a He-atom bounded by fullerene and calculated the
ground state properties. The He-atom with its two electrons and
nucleus centered at $\vec{x}$ has an anti-symmetric excited state
wave function. The method described in this paper allowed us to
calculate the ground state energy as functions of attractive
potential and the well thickness.

In Fig.~5 we show the ground state energies as a function of
attractive energy, $U_0$ for different $D$ and thickness values.
Similar to the H-atom results, electrons become more immobile with
increasing the shell thickness. Note that in the limit of zero
$U_0$, the ground state energy gets the well-known value of the He
ground state energy which is $-2.903$ ($-78.96 eV$). The ground
state of energy decreases after the threshold $U_0\approx-1.9 $
value for the centered case and the threshold value is difference
from the H-atom problem. Numerical values of ground state energies
are given in the Table IV.

Finally, in Fig.~6 we show electron distributions in the $x-z$ plane
when nucleus remains at origin for a high value of $U_0$. Clearly,
two electrons prefer to stay far from each other. Moreover by
increasing the value of $U_0$, the probability of finding each
electron near the shell increases.

\section{Concluding remarks}

In this work we have numerically calculated the ground state
energies and electron distributions of a simple atom confined by
C$_{60}$ fullerene within the diffusion Monte Carlo method. We
modeled the C$_{60}$ fullerene with attractive confining well and
obtained the physical quantities of one or two electrons of the atom
endohedrally. Our results for a hydrogen atom located at the origin
is in excellent agreement with exact calculations. Within the DMC
method we obtained the ground state energy for off-centered H- and
He-atoms for different confining well potentials. We described the
electrons distributions for endohedral case as a function of nucleus
positions and the confining well potential as well. Moreover,
electron distributions are more influenced by changing the nucleus
position. We believe that for an accurate quantitative calculations
of off-centered effects, the real physical model for C$_{60}$
molecule should be deformed in the presence of endohedral atoms and
deviation from spherical symmetry should be considered.

\begin{acknowledgments}
We would like to thank A. Namiranian, M. Fouladvand and N. Nafari
for fruitful discussions.
\end{acknowledgments}
\appendix
\section{Details on explicit calculation of the time step }

The time step can be selected according to the criterion that the
systematic error associated with the use of a finite $\Delta \tau$
is less than the statistical error. To achieve the desired variance
in energy, the value of $N_0$ and run time are chosen sufficiently
large. In these simulations we chose $N_0$ and the run time in the
range of 4000-6000. In Fig~7. we show the energy and its standard
deviation as a function of $\Delta \tau$ for free He-atom. As it is
clear from the figure, the range $[0.05-0.1]$ should be the
acceptable range of the time step, $\Delta \tau$.

\newpage
\begin{figure}
\begin{center}
\includegraphics[scale=0.9]{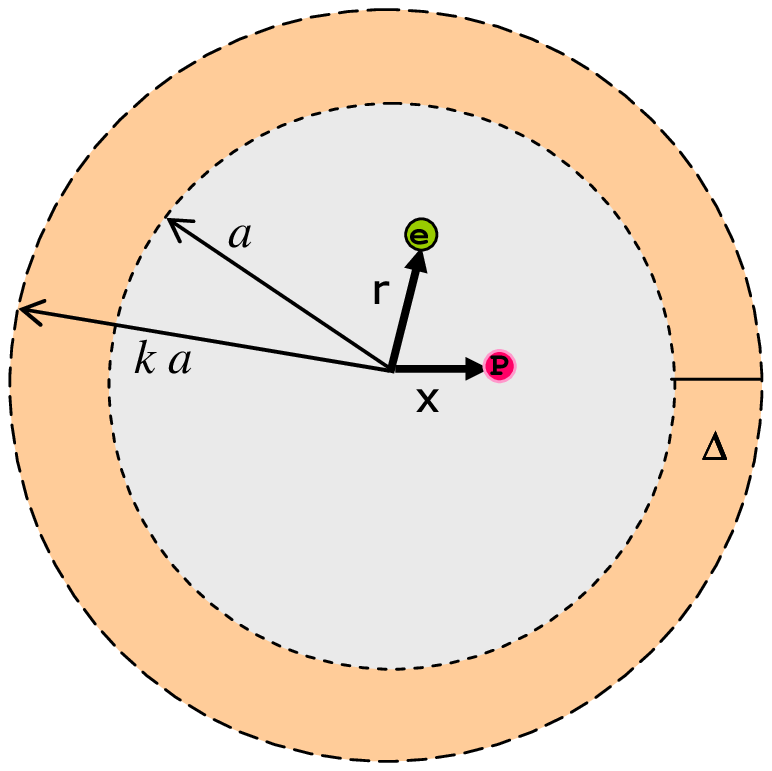}
\caption{(Color online) The schematic representation of H-atom
inside an attractive shell for modeling the H-atom endohedrally
confined inside C$_{60}$ molecules. }
\end{center}
\end{figure}
\newpage
\begin{figure}
\begin{center}
\includegraphics[scale=0.9]{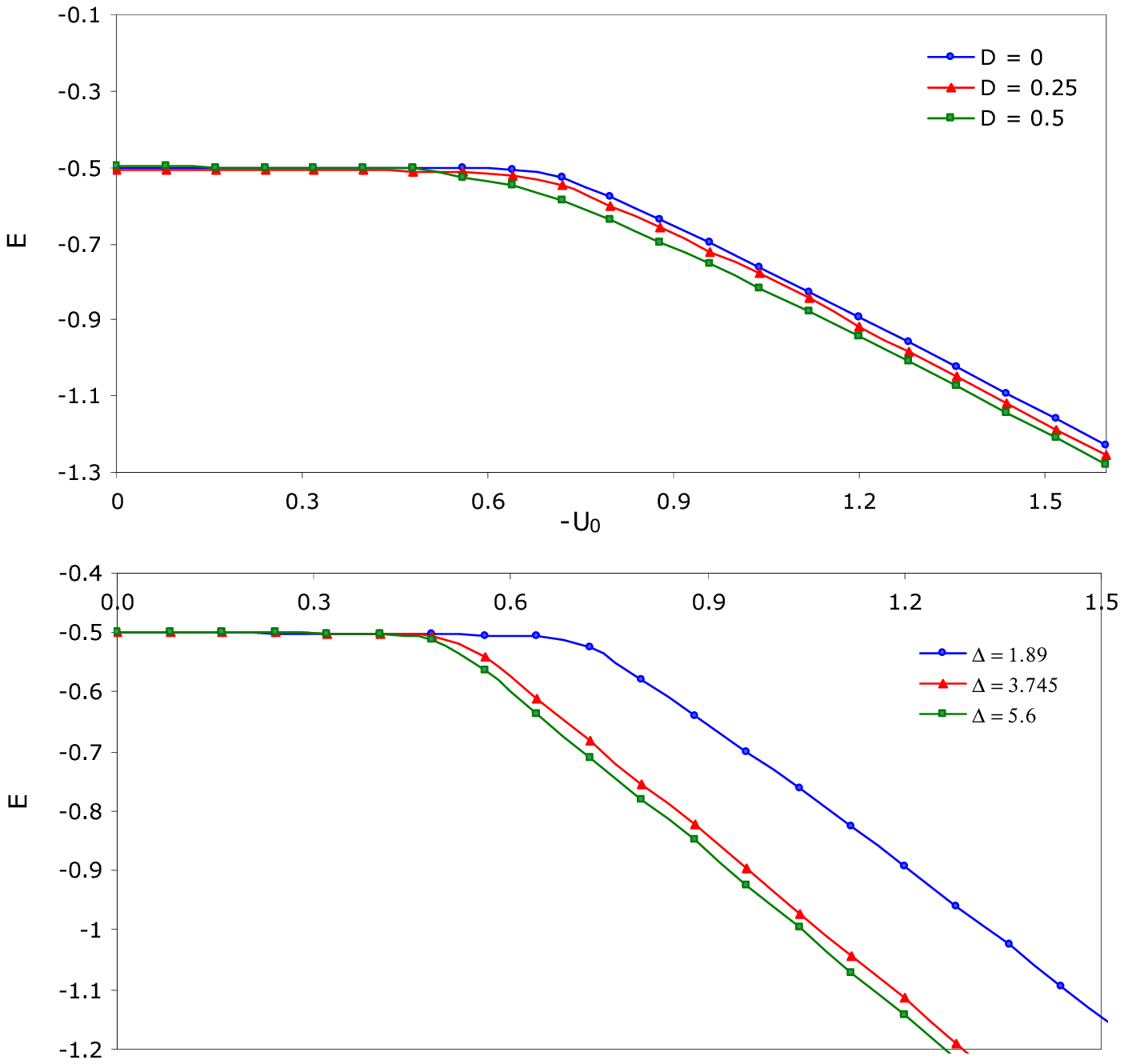}\
\caption{(Color online) Top: Ground state energy for H-atom located
at position $\vec{x}=(0,0,Da)$ as a function of $U_0$ in units of
Hartree. Bottom: Ground state energy of H-atom located at origin as
a function of $U_0$ in units of Hartree for different thickness
values.  Here $a$ is $5.75$.}
\end{center}
\end{figure}
\newpage
\begin{figure}
\begin{center}
\includegraphics[scale=0.9]{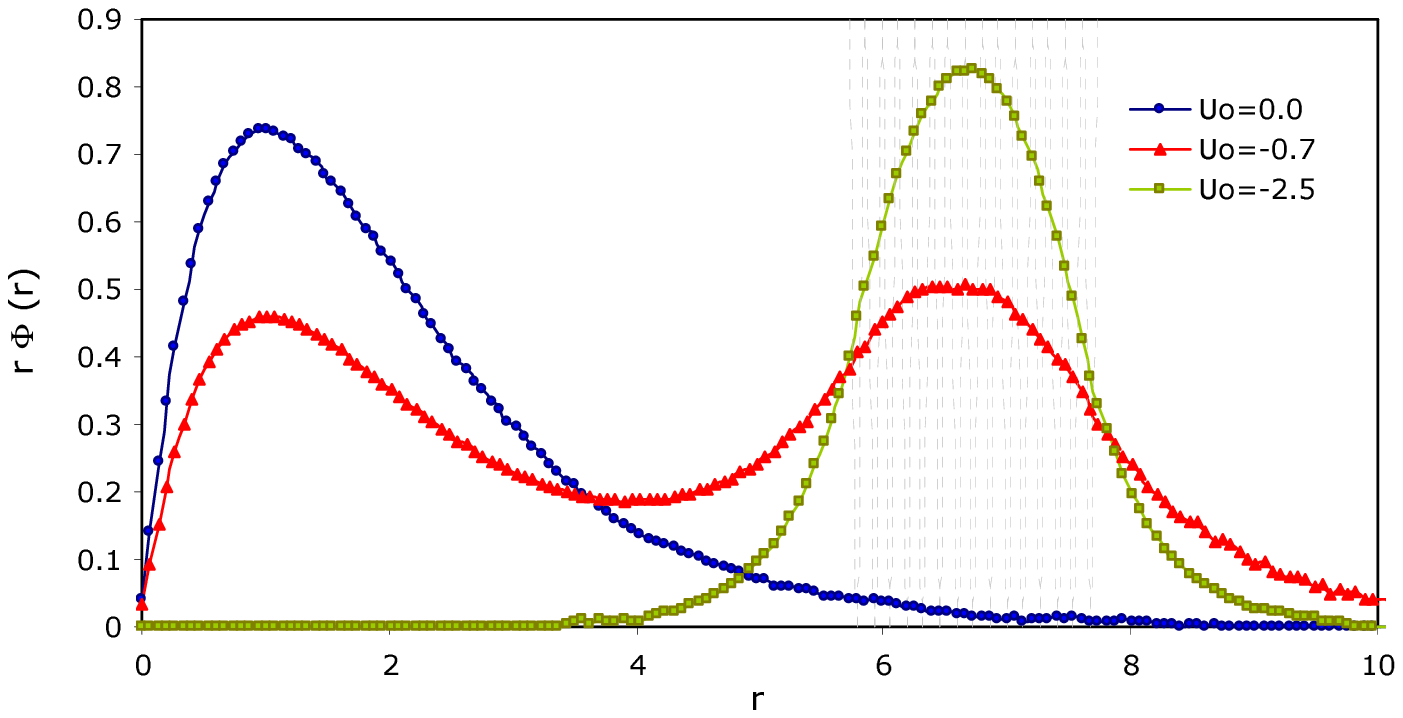}
\caption{(Color online) Radial electron wave function multiply to
position, $r\phi(r)$ as a function of $r$ for H-atom located at
$D=0$ for different $U_0$ values. Dashed vertical lines refer to the
shell region. }
\end{center}
\end{figure}
\newpage
\begin{figure}
\includegraphics[width=10cm]{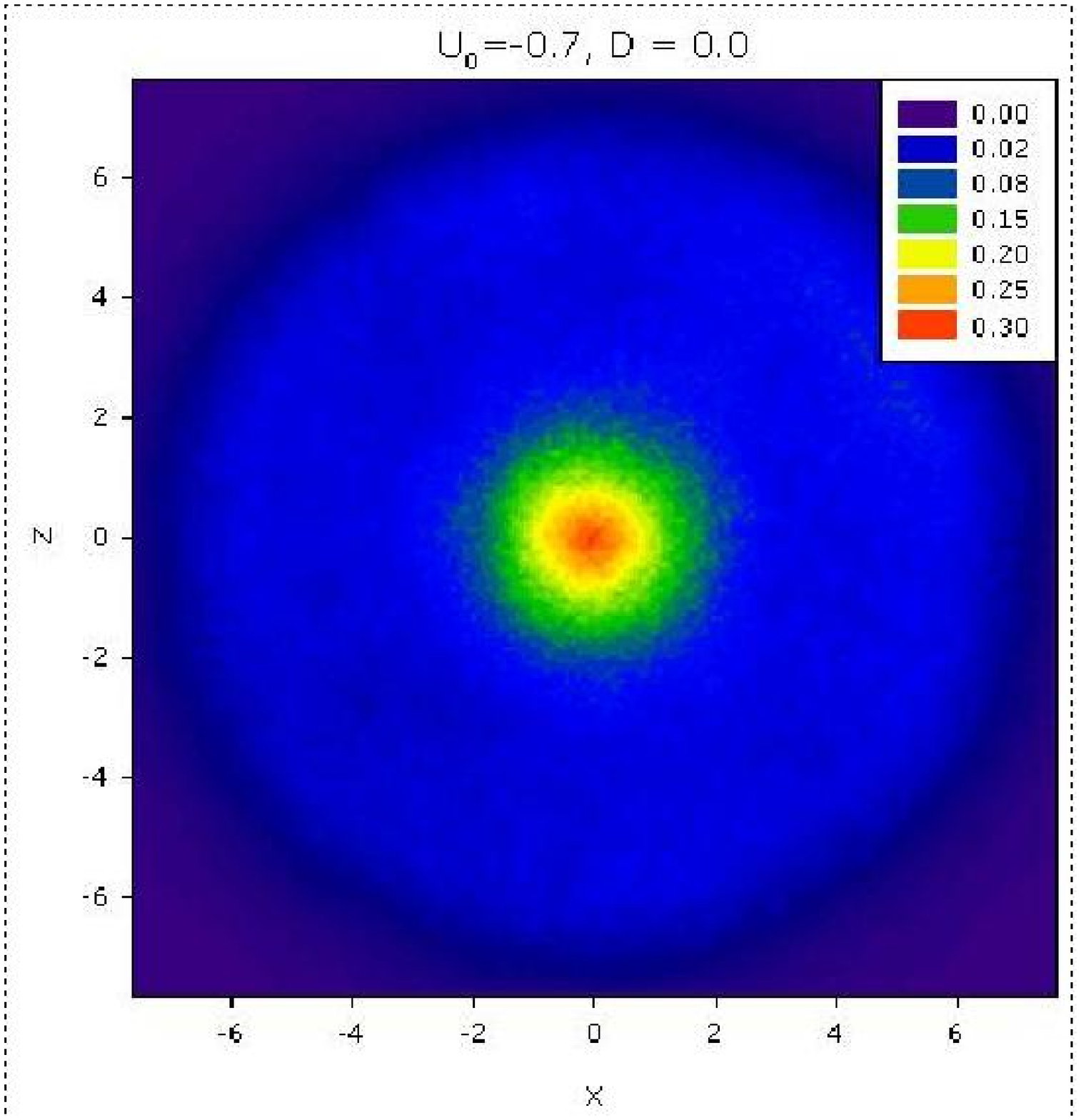}\\
\includegraphics[width=10cm]{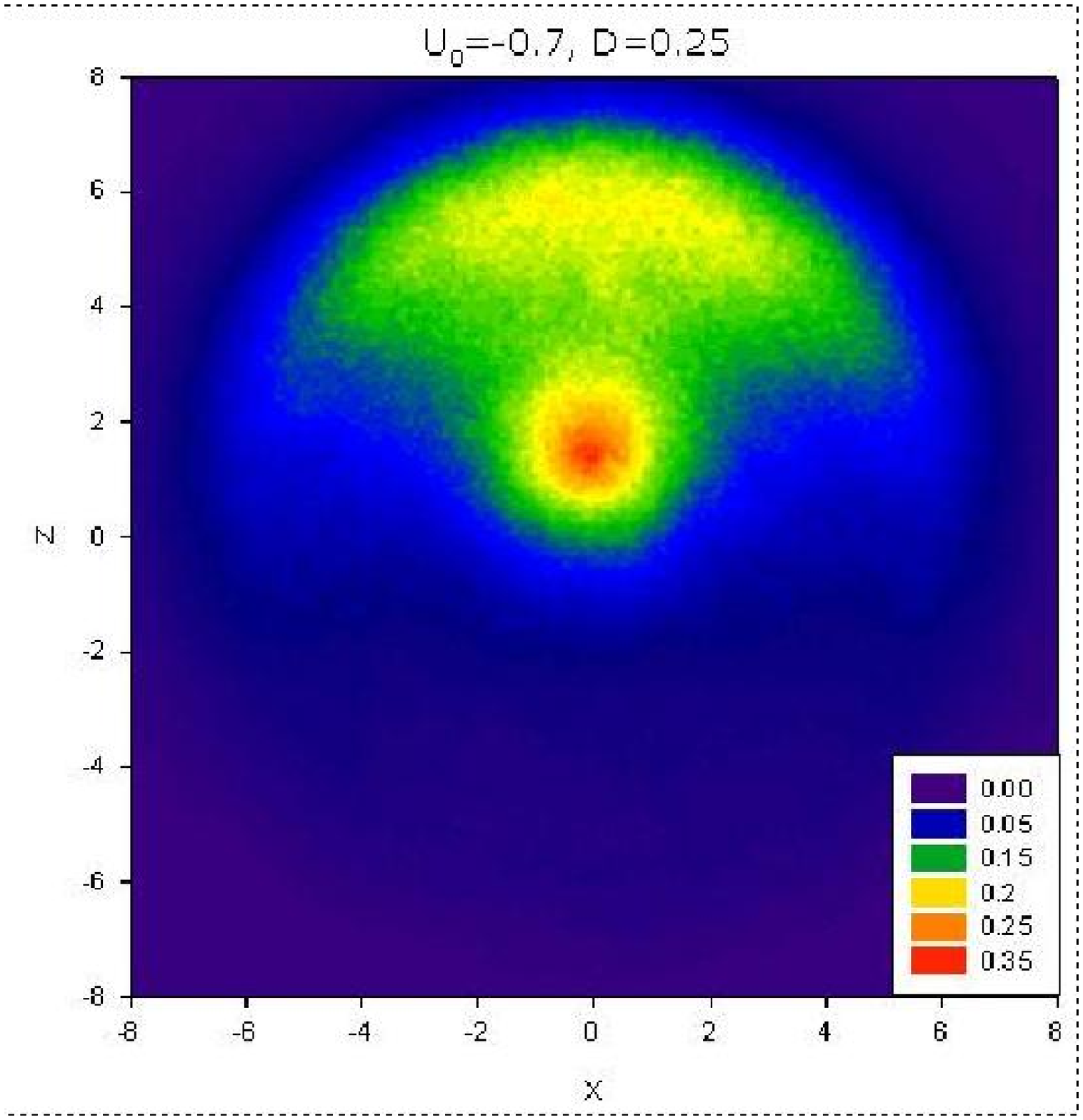}
\caption{(Color online) Electron distributions of H-atom in $x-z$
plane for $U_0=-0.7$ where $D=0.0$ (top) and $D=0.25$ (bottom).
Legends denote the number of normalized electron density.}
\end{figure}
\newpage
\begin{figure}
\begin{center}
\includegraphics[scale=0.8]{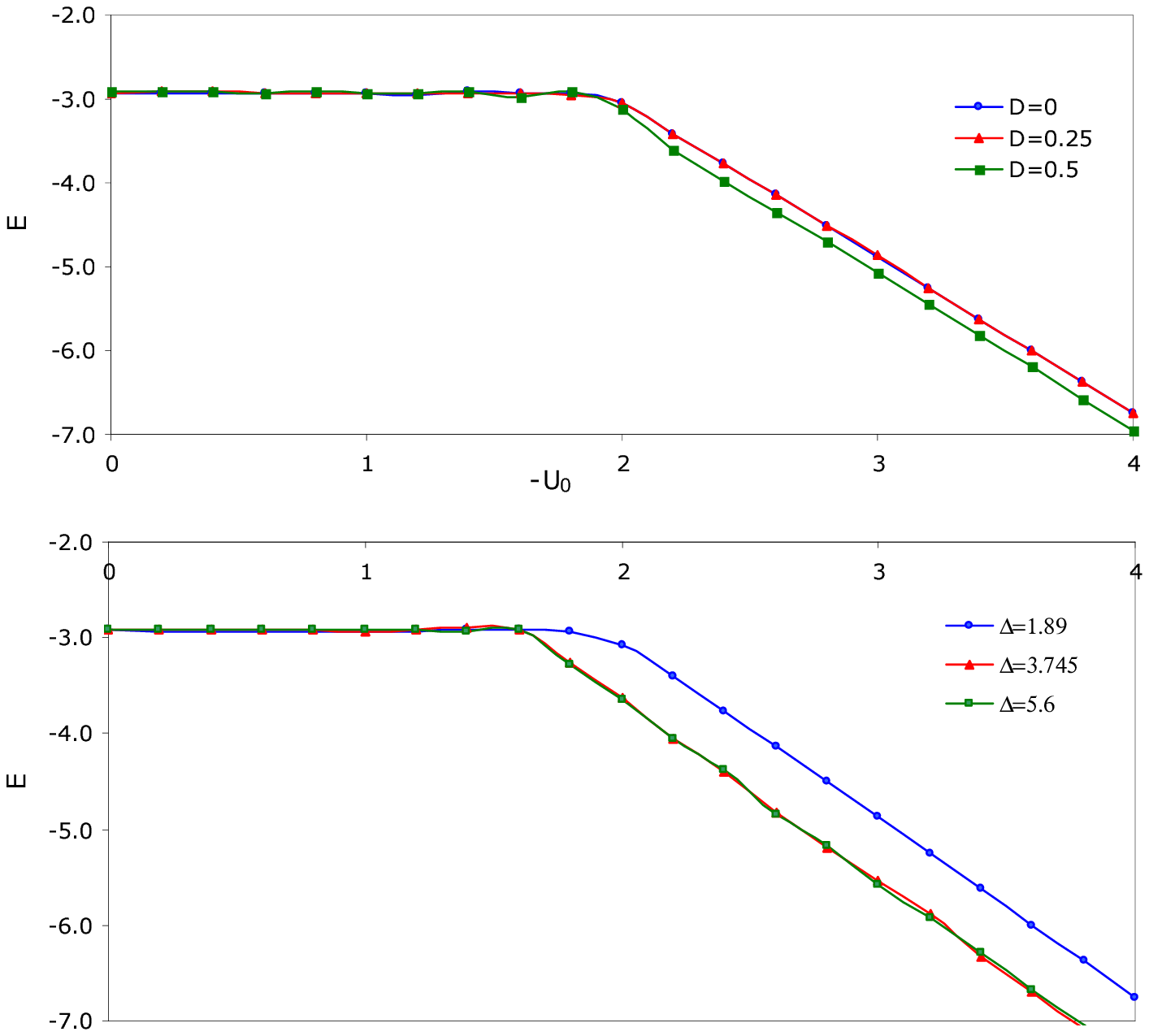}
\caption{(Color online) Top: Ground state energy of He-atom as a
function of $U_0$ in units of Hartree. The nuclear is located at
$\vec{x}=(0,0,Da)$ for $D=0, 0.25$ and $0.5$. Bottom: Ground state
energy of He-atom as a function of $U_0$ in units of Hartree for
different thickness values.  Here $a$ is $5.75$. }
\end{center}
\end{figure}
\newpage
\begin{figure}
\includegraphics[width=10cm]{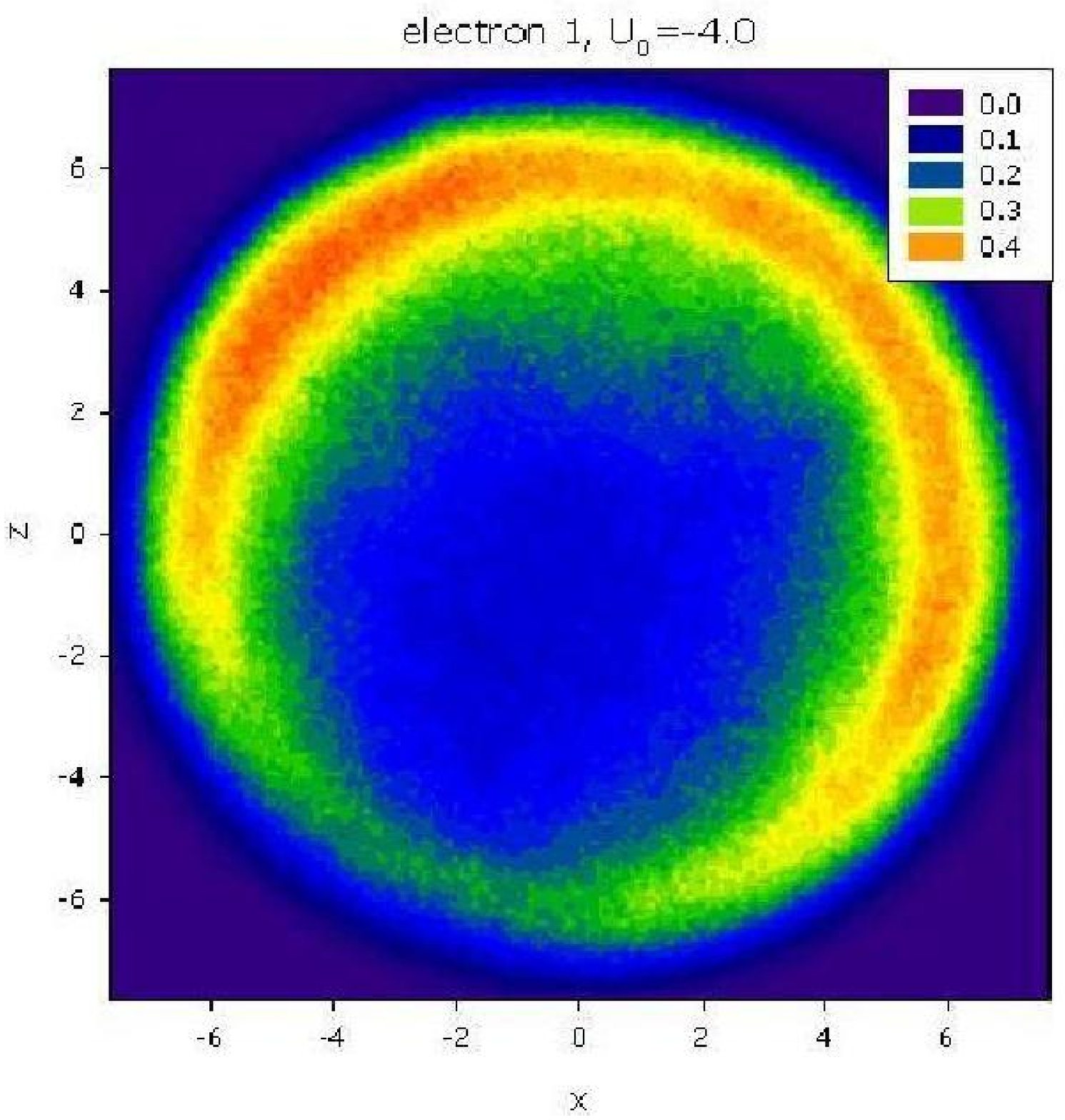}\\
\includegraphics[width=10cm]{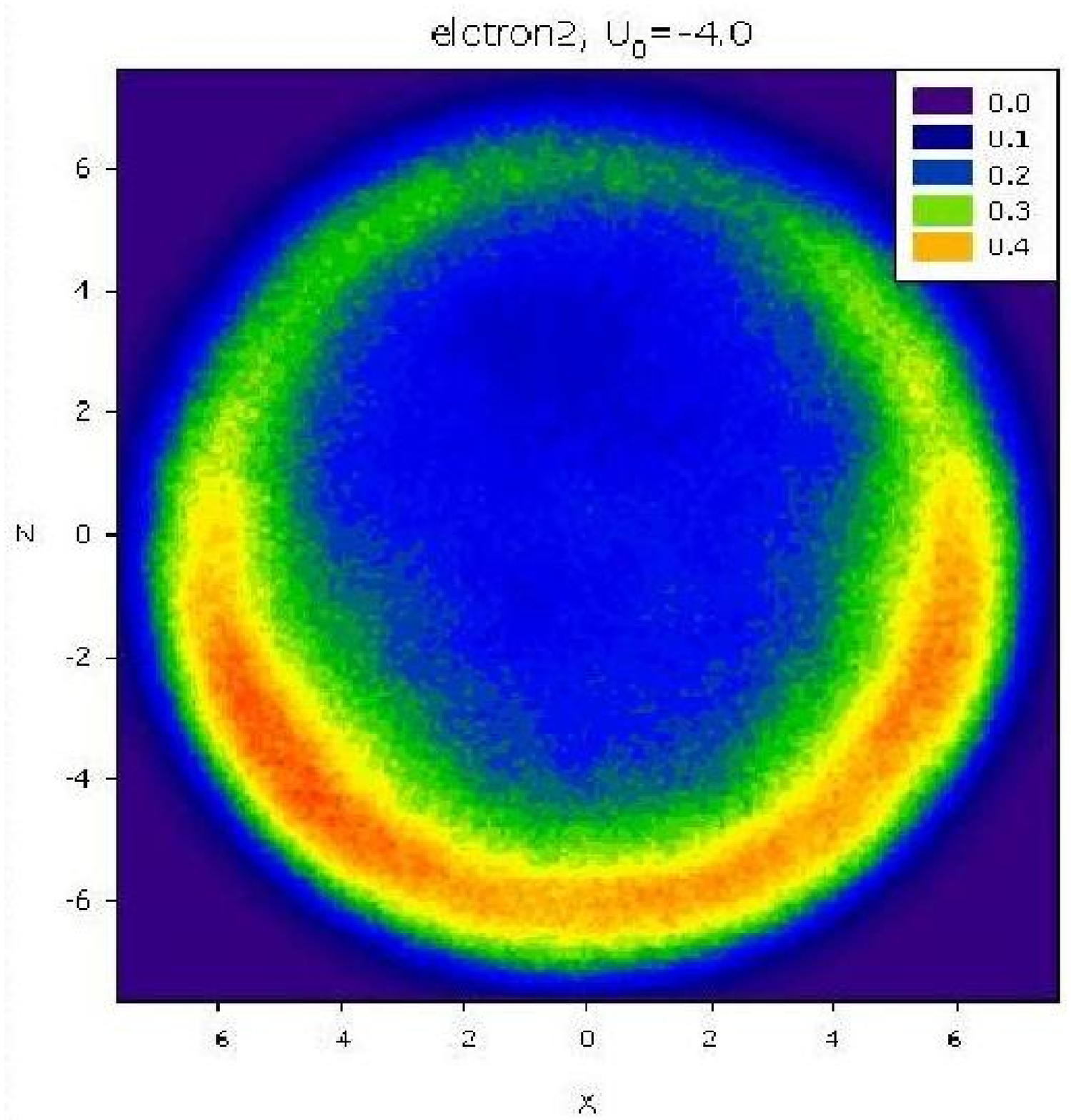}
\\ \caption{(Color online) Electron distributions in $x-z$ plane
for $U_0=-4.0$ for two electrons of He-atom confined at $D=0.0$.
Legends denote the number of normalized electron density.}
\end{figure}
\newpage
\begin{figure}
\begin{center}
\includegraphics[scale=0.9]{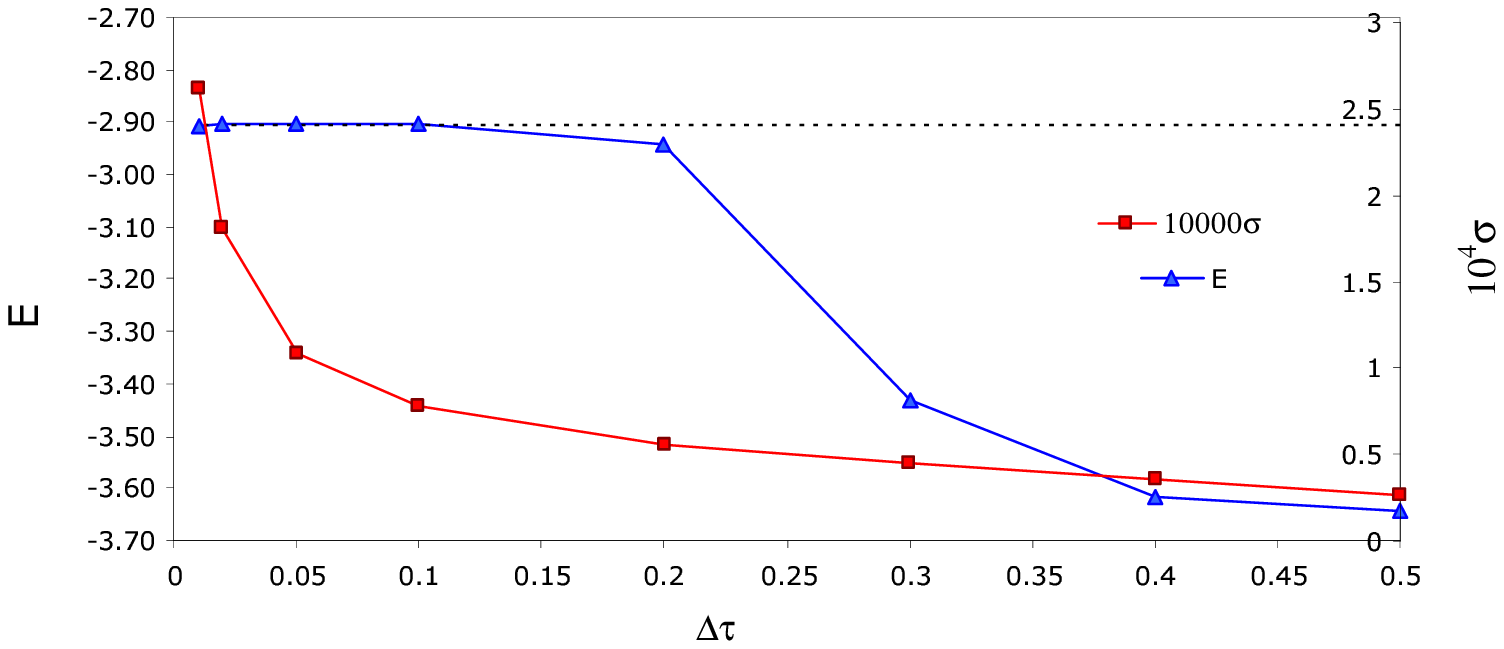}
\caption{(Color online) Energy (triangle symbol) and standard
deviation (square symbol) as a function of time step for free
He-atom.}
\end{center}
\end{figure}
\begin{table}[t]
\begin{tabular}{|c|c|c|c|}
\hline
-$U_0$  & D=0.0  & D=0.25&  D=0.5 \\
\hline
$0.00$ &  $-0.5000$ &  $-0.5010$ &  $-0.5035$ \\
$0.08$ &  $-0.5011$ &  $-0.5046$ &  $-0.5060$ \\
$0.16$ &  $-0.5017$ &  $-0.5050$ &  $-0.5069$ \\
$0.24$ &  $-0.5019$ &  $-0.5055$ &  $-0.5073$ \\
$0.32$ & $-0.5023$ &  $-0.5057$ &  $-0.5083$ \\
$0.40$ &  $-0.5027$ &  $-0.5072$ &  $-0.5090$ \\
$0.48$ &  $-0.5032$ &  $-0.5099$ &  $-0.5120$ \\
$0.56$ &  $-0.5039$ &  $-0.5124$ &  $-0.5251$ \\
$0.64$ &  $-0.5053$ &  $-0.5200$ &  $-0.5461$ \\
$0.72$ &  $-0.5255$ &  $-0.5490$ &  $-0.5866$ \\
$0.80$ &  $-0.5789$ &  $-0.6035$ &  $-0.6397$ \\
$0.88$ &  $-0.6388$ &  $-0.6563$ &  $-0.6970$ \\
$0.96$ &  $-0.6998$ &  $-0.7211$ &  $-0.7549$ \\
$1.04$ &  $-0.7628$ &  $-0.7796$ &  $-0.8191$ \\
$1.12$ &  $-0.8271$ &  $-0.8413$ &  $-0.8806$ \\
$1.20$ &  $-0.8934$ &  $-0.9196$ &  $-0.9455$ \\
$1.28$ &  $-0.9585$ &  $-0.9848$ &  $-1.0104$ \\
$1.36$ &  $-1.0259$ &  $-1.0508$ &  $-1.0756$ \\
$1.44$ &   $-1.0938$ & $-1.1176$ & $-1.1438$ \\
$1.52$ &  $-1.1619$ &  $-1.1886$ &  $-1.2112$ \\
$1.60$ &  $-1.2311$ &  $-1.2562$ &  $-1.2800$ \\
 \hline
\end{tabular}
\caption{Ground state energy for H-atom for different $D$ values.
The statistical error bar for GDMC is about $\pm 10^{-4}$.}
\label{ResultList}
\end{table}

\begin{table}[t]
\begin{tabular}{|c|c|c|c|}
\hline
-$U_0$ & $\Delta=1.89$ & $\Delta=3.745$ & $\Delta=5.6 $ \\
\hline
$0.00$ & $-0.5000$ & $ -0.5000$ & $ -0.5000$  \\
$0.08$ & $-0.5002$ & $ -0.5003$ & $ -0.5005$  \\
$0.16$ & $-0.5005$ & $ -0.5007$ & $ -0.5007$ \\
$0.24$ & $-0.5009$ & $ -0.5011$ & $ -0.5014$\\
$0.32$ & $-0.5010$ & $ -0.5014$ & $ -0.5025$  \\
$0.40$ & $-0.5011$ & $ -0.5015$ & $ -0.5027$  \\
$0.48$ & $-0.5012$ & $ -0.5062$ & $ -0.5131$  \\
$0.56$ & $-0.5056$ & $ -0.5409$ & $ -0.5639$  \\
$0.64$ & $-0.5061$ & $ -0.6100$ & $ -0.6371$  \\
$0.72$ & $-0.5259$ & $ -0.6824$ & $ -0.7095$  \\
$0.80$ & $-0.5789$ & $ -0.7544$ & $ -0.7800$ \\
$0.88$ & $-0.6399$ & $ -0.8238$ & $ -0.8480$  \\
$0.96$ & $-0.7001$ & $ -0.8965$ & $ -0.9234$  \\
$1.04$ & $-0.7622$ & $ -0.9718$ & $ -0.9941$  \\
$1.12$ & $-0.8271$ & $ -1.0437$ & $ -1.0707$  \\
$1.20$ & $-0.8929$ & $ -1.1128$ & $ -1.1423$ \\
$1.28$ & $-0.9584$ & $ -1.1907$ & $ -1.2157$  \\
$1.36$ & $-1.0242$ & $ -1.2637$ & $ -1.2857$  \\
$1.44$ & $-1.0934$ & $ -1.3390$ & $ -1.3581$  \\
$1.52$ & $-1.1612$ & $ -1.4142$ & $ -1.4311$  \\
$1.60$ & $-1.2297$ & $ -1.4819$ & $ -1.5016$  \\
 \hline
\end{tabular}
\caption{Ground state energy for H-atom for different thickness
values.} \label{ResultList}
\end{table}
\begin{table}[t]
\begin{tabular}{|c|c|c|c|}
\hline
-$U_0$ & D=0.0 & D=0.25  & D=0.5  \\
\hline
$0.00$ &  $-2.9031$ &  $-2.9036$ &  $-2.9038$\\
$0.20$ &  $-2.9033$ &  $-2.9045$ &  $-2.9103$\\
$0.40$ &  $-2.9040$ &  $-2.9057$ &  $-2.9116$\\
$0.60$ &  $-2.9044$ &  $-2.9063$ &  $-2.9138$\\
$0.80$ &  $-2.9055$ &  $-2.9077$ &  $-2.9144$\\
$1.00$ &  $-2.9061$ &  $-2.9088$ &  $-2.9152$\\
$1.20$ & $-2.9064$ &  $-2.9145$ &  $-2.9169$\\
$1.40$ &  $-2.9073$ &  $-2.9188$ &  $-2.9176$\\
$1.60$ &  $-2.9081$ &  $-2.9231$ &  $-2.9276$\\
$1.80$ &  $-2.9087$ &  $-2.9424$ &  $-2.9445$\\
$2.00$ &  $-3.0523$ &  $-3.0560$ &  $-3.0653$\\
$2.20$ &  $-3.4101$ &  $-3.4148$ &  $-3.6131$\\
$2.40$ &  $-3.7751$ &  $-3.7756$ &  $-3.9799$\\
$2.60$ &  $-4.1364$ &  $-4.1441$ &  $-4.3421$\\
$2.80$ &  $-4.5043$ &  $-4.5053$ &  $-4.7079$\\
$3.00$ &  $-4.8758$ &  $-4.8705$ &  $-5.0756$\\
$3.20$ &  $-5.2496$ &  $-5.2502$ &  $-5.4505$\\
$3.40$ &  $-5.6228$ &  $-5.6228$ &  $-5.8193$\\
$3.60$ &  $-5.9986$ &  $-5.9976$ &  $-6.1959$\\
$3.80$ &  $-6.3742$ &  $-6.3767$ &  $-6.5741$\\
$4.00$ &  $-6.7538$ &  $-6.7494$ &  $-6.9580$\\
 \hline
\end{tabular}
\caption{Ground state energy for He-atom for different $D$ values.}
\label{ResultList}
\end{table}

\begin{table}[t]
\begin{tabular}{|c|c|c|c|}
\hline
-$U_0$ &$\Delta=1.89$ &$\Delta=3.745$ &$\Delta=5.6$ \\
\hline
$0.00$ & $-2.9031$ & $-2.9032$ & $-2.9032$ \\
$0.20$ & $-2.9252$ & $-2.9258$ & $-2.9263$ \\
$0.40$ & $-2.9260$ & $-2.9261$ & $-2.9265$ \\
$0.60$ & $-2.9333$ & $-2.9363$ & $-2.9372$ \\
$0.80$ & $-2.9344$ & $-2.9372$ & $-2.9394$ \\
$1.00$ & $-2.9388$ & $-2.9396$ & $-2.9429$ \\
$1.20$ & $-2.9435$ & $-2.9455$ & $-2.9468$ \\
$1.40$ & $-2.9453$ & $-2.9469$ & $-2.9479$ \\
$1.60$ & $-2.9559$ & $-2.9614$ & $-2.9738$ \\
$1.80$ & $-2.9657$ & $-3.2565$ & $-3.2902$ \\
$2.00$ & $-3.0823$ & $-3.6351$ & $-3.6563$ \\
$2.20$ & $-3.4101$ & $-4.0522$ & $-4.0453$ \\
$2.40$ & $-3.7751$ & $-4.4060$ & $-4.3871$ \\
$2.60$ & $-4.1364$ & $-4.8250$ & $-4.8418$ \\
$2.80$ & $-4.5043$ & $-5.1915$ & $-5.1753$ \\
$3.00$ & $-4.8758$ & $-5.5377$ & $-5.5693$ \\
$3.20$ & $-5.2496$ & $-5.8922$ & $-5.9155$ \\
$3.40$ & $-5.6228$ & $-6.3206$ & $-6.2839$ \\
$3.60$ & $-5.9986$ & $-6.6892$ & $-6.6714$ \\
$3.80$ & $-6.3742$ & $-7.0863$ & $-7.0453$ \\
$4.00$ & $-6.7538$ & $-7.4316$ & $-7.4289$ \\
 \hline
\end{tabular}
\caption{Ground state energy for He-atom for different thickness
values.} \label{ResultList}
\end{table}
\newpage
\end{document}